# A Memory Aware High Level Synthesis Tool

Gwenolé Corre, Eric Senn, Nathalie Julien and Eric Martin
L.E.S.T.E.R., University of South-Brittany 56321 Lorient cedex, France
gwenole.corre@univ-ubs.fr

## Abstract

*We introduce a new approach to take into account the memory architecture and the memory mapping in High-Level Synthesis for data intensive applications. We formalize the memory mapping as a set of constraints for the synthesis, and defined a Memory Constraint Graph and an accessibility criterion to be used in the scheduling step. We use a memory mapping file to include those memory constraints in our HLS tool GAUT. It is possible, with the help of GAUT, to explore a wide range of solutions, and to reach a good tradeoff between time, power-consumption, and area.*

## 1. Introduction

To tackle the complexity of memory design, we consider as essential to take into account memory accesses directly during the behavioral synthesis, assuming that a reasonable trade-off between the design time and the quality of the results is reached.

In this paper, we propose a new and simple technique to take into account the memory mapping in the high level synthesis. Indeed, our aim is to produce a simple algorithm to achieve the synthesis of even complex designs in a reasonable time. We focus on the definition of a memory mapping file that is used in the synthesis process. We introduce an original scheduling in the synthesis flow, to obtain an optimized RTL design, and present a formalism to resolve scheduling under memory constraint. Our methodology was implemented in our HLS tools GAUT[1]. Several syntheses were performed that exhibit the interest of our approach. Experimental results are discussed in conclusion.

## 2. Memory integration

### 2.1. Memory aware synthesis

We introduce memory synthesis in the standard HLS design flow.

The difference between the standard and the memory aware design flow is illustrated on Fig 1. A Signal Flow Graph (SFG) is first generated from the algorithmic specification. In the new approach, this SFG is parsed and a memory table is created. This memory table is then completed by the designer who can select the variable implementation (memory or register) and place the variable in the memory hierarchy (which bank). The resulting table is the memory mapping that will be used in the synthesis. In the standard flow, the processing unit is synthesized without any knowledge on the memory mapping. The memory architecture is designed afterward and a lot of optimization opportunities are definitely lost.

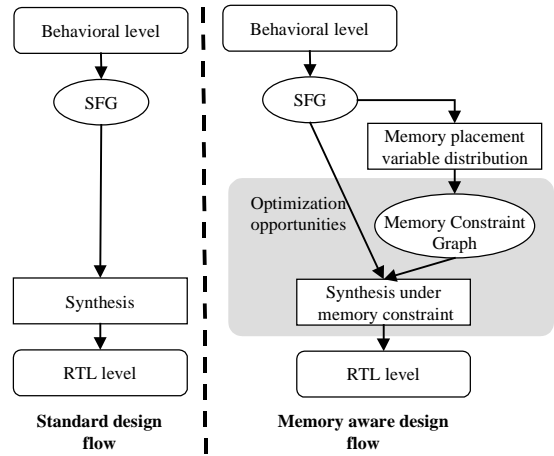

**Figure 1 : Standard and memory aware design flows**

The memory mapping file contains information about every data structure in the algorithm (mainly arrays in DSP applications) and its allocation in memory (bank number and physical address). Scalars can also be defined. This memory table represents all data vertices extracted from a SFG. This data distribution can be static or dynamic. In the case of a static placement, the data stay at the same place during the whole execution. If the placement is dynamic, data can be transferred between different levels in the memory hierarchy. Thus, several data can share the same location in the circuit memory. The memory mapping file explicitly describes the data transfers to occur during the algorithm execution. Direct Memory Address (DMA) directives will be added to the code to achieve these transfers. Definition of the memory architecture is performed in the first step of the overall design flow. To achieve this task, advanced compilers such as Rice HPF compiler, Illinois Polaris or Stanford SUIF could be used[2]. Indeed, these compilers automatically perform data distribution across banks, determine which access goes to which bank, and then schedule to avoid bank conflicts. The Data Transfer and Storage Exploration (DTSE) method from IMEC[3] and the associated tools (ATOMIUM, ADOPT) are also a good mean to determine a convenient data mapping.

### 2.2. Signal Flow graph

The input of our HLS tool is an algorithmic description specifies the circuit's functionality at the behavioral level, disregarding any potential implementation solutions. This initial description is compiled in order to obtain an intermediate representation, the Signal Flow Graph (SFG). The difference between a Signal Flow Graph and Data Flow Graph resides in the introduction of delay operators ($z^{-1}$). These operators are necessary to express the use of data whose value was computed in a preceding iteration of the algorithm. A vertex represents one of the following operations: arithmetic, logical, data or delay. An edge $E_{i,j} = (v_i, v_j)$ represents a data dependence between operations $v_i$ and $v_j$ such as for any iteration of the SFG, operation $v_i$ must start its execution before that of $v_j$. For the data dependencies, the execution of $v_j$ can start only after the completion of operation $v_i$.

### 2.3. Memory Constraint Graph

As outlined in section 2.1, all data vertices are extracted from the SFG to construct the memory table. The designer can choose the data to be placed in memory and defines a memory mapping. For every memory in the memory table, we construct a weighted Memory Constraint Graph (MCG). It represents conflicts and scheduling possibilities between all nodes placed in this memory. The MCG is constructed from the SFG and the memory mapping file. It will be used during the scheduling step of the synthesis. Memory Constraint Graphs are used during the scheduling process to determine the accessibility criterion and the time of every memory access.

### 2.4. Scheduling under memory constraint

The classical list scheduling algorithm relies on heuristics in which ready operations (operations to be scheduled) are listed by priority order. In our tool, an early scheduling is performed. In this scheduling, the priority function depends on the mobility criterion. This mobility is computed, for each cycle, as the difference, in number of cycles, between the current cycle and the operation deadline. Whenever two ready operations need to access the same resource (this is a so called resource conflict), the operation with the lower mobility has the highest priority and is scheduled. The other is postponed. To perform a scheduling under memory constraint, we introduce fictive memory access operators and add an accessibility criterion based on the MCG. A memory has as much access operators as access ports. The memory is declared accessible if one of its fictive memory access operators is idle. Several operations can try to access the same memory in the same cycle; accessibility is used to determine which operations are really executable. Fictive memory access operators are represented by tokens on the MCG. There are as many tokens in the MCG as ports (R/W) in the memory. These tokens are used to compute the accessibility of the memory. The list of ready operations is still organized according to the mobility criterion, but all the operations that do not match the accessibility condition are removed from this list. To schedule an operation that involves an access to the memory, we check if the data is not in a busy memory bank. If a memory bank is not available, every operation that needs to access this memory will not be scheduled, no matter its priority level.

## 3. Conclusion

Several experiences were made, that exhibits the interest of this approach. Memory aware synthesis and GAUT appear very efficient for exploring the design space and for balancing optimizations between the processing unit and the memory unit. It permits to determine the best memory architecture, i.e. the best number of memory banks, as well as the best memory mapping, to meet the application constraints, and to finally reach a reasonable tradeoff between time, power consumption, and area.